\documentclass{PoS}

\title{$B \to \pi\ell\nu$ form factors and $|V_{ub}|$ with M\"obius domain
wall fermions}

\ShortTitle{$B \to \pi\ell\nu$ form factors and $|V_{ub}|$ with M\"obius domain
wall fermions}

\author{B. Colquhoun$^{a}$, S. Hashimoto$^{b,c}$, T. Kaneko$^{b,c}$, and \speaker{J. Koponen}$^{,b}$\\
  \llap{$^{a}$}Department of Physics and Astronomy, York University, Toronto, ON, M3J 1P3, Canada\\
  \llap{$^{b}$}High Energy Accelerator Research Organization (KEK), Ibaraki 305-0801, Japan\\
  \llap{$^{c}$}School of High Energy Accelerator Science, SOKENDAI (The Graduate University for
  Advanced Studies), Ibaraki 305-0801, Japan\\
        E-mail: \email{jonna.koponen@kek.jp}}

\abstract{
We report on a calculation of form factors
for the semileptonic decay of $B$ meson to pion on
$2+1$-flavour lattices with lattice spacings from 0.080 fm down to 0.044
fm. Using the M\"obius domain wall fermion action for both sea and
valence  quarks, we simulate pions with masses down to 225 MeV.
By utilizing a range of heavy quark masses up to 2.44 times the
mass of the charm quark we extrapolate to the physical $b$ quark
mass. We discuss the dependence of the form factors on the pion
mass, heavy quark mass, lattice spacing and the momentum-transfer. 
We extract the CKM matrix element $|V_{ub}|$ through a simultaneous 
fit with the $B \to \pi\ell\nu$ differential branching fractions provided 
by the  Belle and BaBar collaborations after a chiral-continuum 
and physical $b$ quark extrapolations of our lattice data.
}

\FullConference{37th International Symposium on Lattice Field Theory - Lattice2019\\
		16-22 June 2019\\
		Wuhan, China}

\usepackage{amsmath}

\begin{document}

\section{Introduction}

The semileptonic process $B \to \pi\ell\nu$ may be used to extract the element $|V_{ub}|$
of the Cabibbo--Kobayashi--Maskawa matrix. Here we report on our lattice QCD study of this decay,
which forms a part of a larger series of studies of heavy quark processes, including other
exclusive decays like $B \to D^{(\ast)}$ \cite{Takashi_PoSLATT2019} and inclusive decays
\cite{inclusive}. We use the M\"obius domain-wall fermion action \cite{MobiusDW} for
all quarks, which has the advantage of including all relativistic effects for the heavy quarks,
but necessitates extrapolating to physical $m_b$ from lower heavy quark masses, $m_h$.
Preliminary results have been reported in \cite{Brian_PoSLATT2018}.

The CKM matrix element can be related to the (experimental) differential decay rate by
\begin{equation}
\frac{\mathrm{d}\Gamma (B \to\pi\ell\nu)}{\mathrm{d}q^2}
= \frac{G_F^2|V_{ub}|^2}{24\pi^3}|k_\pi|^3|f_+(q^2)|^2.
\label{eq:decayrate}
\end{equation}
Thus calculating the form factor $f_+(q^2)$ from lattice QCD allows us to extract
$|V_{ub}|$. Here $k_\pi$ is the pion four momentum in the $B$ meson rest frame and
$q^\mu=p_B^\mu-k_\pi^\mu$ is the momentum transfer. $p_B$ denotes the four momentum
of the $B$ meson.

\section{Form factors}

For a pseudoscalar to pseudoscalar decay, the vector matrix element can be written as
\begin{equation}
  \langle \pi(k_\pi)|V^\mu|B(p_B)\rangle = f_+(q^2)\bigg[(p_B+k_\pi)^\mu
    -\frac{M_B^2-M_\pi^2}{q^2}q^2\bigg]+f_0(q^2)\frac{M_B^2-M_\pi^2}{q^2}q^\mu.
\end{equation}
We have two form factors, $f_+(q^2)$ and $f_0(q^2)$. These are both calculable in
lattice QCD, even though $f_0(q^2)$ is not accessible experimentally as it is
suppressed by the small lepton mass.

Useful parametrisation in the context of Heavy Quark Effective Theory (HQET) \cite{HQET} is
\begin{equation}
  \langle \pi(k_\pi)|V^\mu|B(p_B)\rangle = 2\sqrt{M_B}\bigg[f_1(v\cdot k_\pi)v^\mu
    +f_2(v\cdot k_\pi)\frac{k_\pi^\mu}{v\cdot k_\pi}\bigg],
  \label{eq:piVB}
\end{equation}
where $v^\mu = p_B^\mu/M_B$ is the heavy quark velocity and
$E_\pi = v\cdot k_\pi = M_B^2+M_{\pi}^2-q^2/(2M_B)$.
The HQET form factors $f_1(v\cdot k_\pi)$ and $f_2(v\cdot k_\pi)$ stay finite in the
limit of infinitely heavy $b$ quark. Corrections of the form $1/m_h$ are expected
for finite $m_h$.

These two sets of form factors are not independent, and $f_+(q^2)$ and $f_0(q^2)$ can
be written in terms of $f_1(v\cdot k_\pi)$ and $f_2(v\cdot k_\pi)$ as
\begin{align}
  f_+(q^2)=&\sqrt{M_B}\bigg[\frac{f_2(v\cdot k_\pi)}{v\cdot k_\pi}
    +\frac{f_1(v\cdot k_\pi)}{M_B}\bigg] \\
  f_0(q^2)=&\frac{2}{\sqrt{M_B}}\frac{M_B^2}{(M_B^2-M_\pi^2)}\bigg[f_1(v\cdot k_\pi)
    +f_2(v\cdot k_\pi)-\frac{v\cdot k_\pi}{M_B}\bigg(f_1(v\cdot k_\pi)
    +\frac{M_\pi^2}{(v\cdot k_\pi)^2}f_2(v\cdot k_\pi)\bigg)\bigg]. \nonumber
\end{align}
  
\section{Lattice setup}

\begin{figure}
\centering
\includegraphics[width=0.8\textwidth]{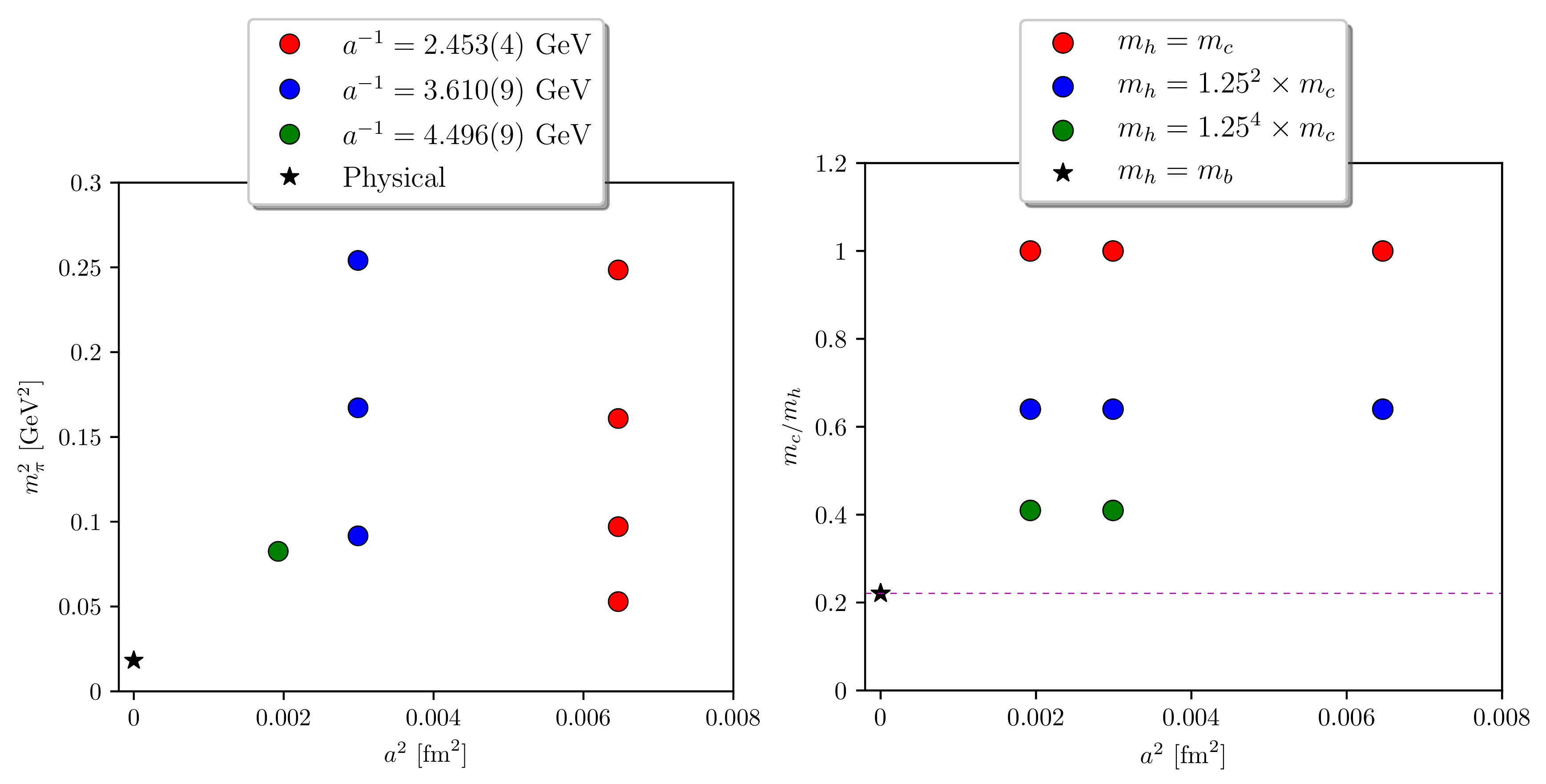}
\caption{On the left: The light sea quark masses used for each lattice
  spacing. The light valence quark masses are the same as the sea quark masses.
  On the right: The heavy valence quark masses used for each lattice spacing.}
\label{fig:lattconfigs}
\end{figure}

We use gauge ensembles generated with $2+1$ flavour M\"obius domain-wall fermions, and the
gauge action is tree-level Symanzik improved. Lattice spacings included in this calculation
are approximately 0.080 fm, 0.055 fm and 0.044 fm, corresponding to $\beta = 4.17$,
$\beta = 4.35$ and $\beta = 4.47$, respectively. Pion masses range from $500$ MeV down to
225 MeV, where we use a larger volume at $\beta = 4.17$ for the lightest pion such that we
maintain $M_{\pi}L > 4$. Heavy quark masses are chosen to be $m_c$, $1.25^2m_c$ and $1.25^4m_c$,
ensuring that $am_h < 0.7$ to avoid large discretization effects from the heavy quark mass.
See Fig.~\ref{fig:lattconfigs} for an illustration of the light and heavy quark masses used
in this study. The plot on the left shows the sea light quark masses used for each lattice
spacing. The valence light quark masses are the same as the sea quark masses. The plot on the
right shows the valence heavy quark masses used for each lattice spacing. Note that the
ensembles and correlators used in the study of the $D \to \pi\ell\nu$
process in \cite{Takashi_EPJC} are a subset of the data used in this study.

The $B$ meson is kept at rest in our calculations while we give pion momenta $p = (0,0,0)$,
$(0,0,1)$, $(0,1,1)$, $(1,1,1)$ in units of $2\pi/L$. We calculate correlators from all permutations
of a given momentum and average these to improve our signal.

We also need to renormalize our vector currents. We have a light quark and a heavy quark at the
current insertion, and we calculate the renormalization factor as $Z_V = \sqrt{Z_{V,hh}Z_{V,ll}}$.
The heavy-heavy renormalisation factor $Z_{V,hh}$ is calculated by demanding that the vector
matrix element $\langle B_s|V|B_s\rangle$ for the heavy current gives 1.
The renormalization
factors for the light current, $Z_{V,ll}$, are from \cite{Zll}. For the lightest heavy quark masses,
i.e. when $am_h = am_c$, we find that it is sufficient to renormalize our currents using results
from the massless coordinate space current correlators as described in \cite{Zll}.
  
\section{Extrapolations}

Choosing the temporal ($\mu=0$) or spatial ($\mu=1,2,3$) vector current in equation \eqref{eq:piVB}
naturally gives the combinations $f_1(v\cdot k_\pi)+f_2(v\cdot k_\pi)$ and $f_2(v\cdot k_\pi)$ of
the form factors. Therefore we use the fit functions
\begin{align}
&f_1(v\cdot k_\pi)+f_2(v\cdot k_\pi)=C_0\bigg(1+\sum_{n=1}^3C_nE^n_\pi+C_4M_\pi^2+\chi_{\textrm{log}}
 +C_5E_\pi M_\pi^2+\frac{C_6}{m_h}\bigg)(1+C_7a^2),  \\
&f_2(v\cdot k_\pi)=D_0\frac{E_\pi}{E_\pi+\Delta_B}\bigg(1+D_1E_\pi
+D_2M_\pi^2+\chi_{\textrm{log}} 
+D_3E_\pi M_\pi^2+\frac{D_4}{m_h}\bigg)(1+D_5a^2) \nonumber
\end{align}
to extrapolate our lattice data to the physical limit: to continuum ($a\to 0$), and to physical
pion and $B$ meson masses.
The extrapolation to physical pion mass is guided by the $M_\pi^2$ and $E_\pi M^2_\pi$ terms and
the chiral logs $\chi_{\textrm{log}}$.
The heavy quark mass dependence is taken to be
of the form $1/m_h$, where we use $m_h = M_{\eta_h}/2$ as a proxy for the heavy quark mass.
($M_{\eta_h}$ is the mass of the pseudoscalar heavy-heavy meson $\eta_h$, i.e. $\eta_b$ at physical heavy
quark mass.) Form factor $f_2(v\cdot k_\pi)$ is expected to have a pole $(E_\pi+\Delta_B)^{-1}$ with
$\Delta_B=M_{B^\ast}-M_B$. Discretisation effects are covered by the $a^2$ terms.

\begin{figure}[tb]
\centering
\includegraphics[width=0.51\textwidth]{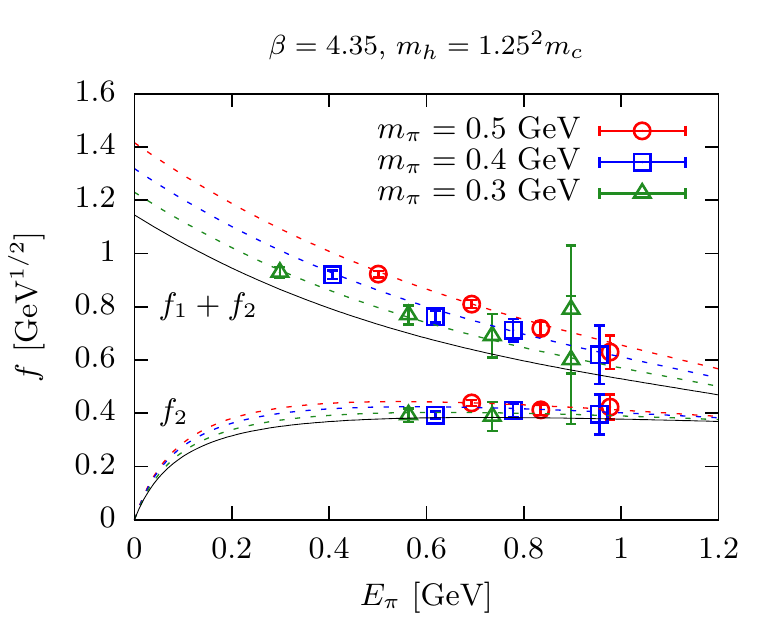}
\hspace*{-4mm}\includegraphics[width=0.51\textwidth]{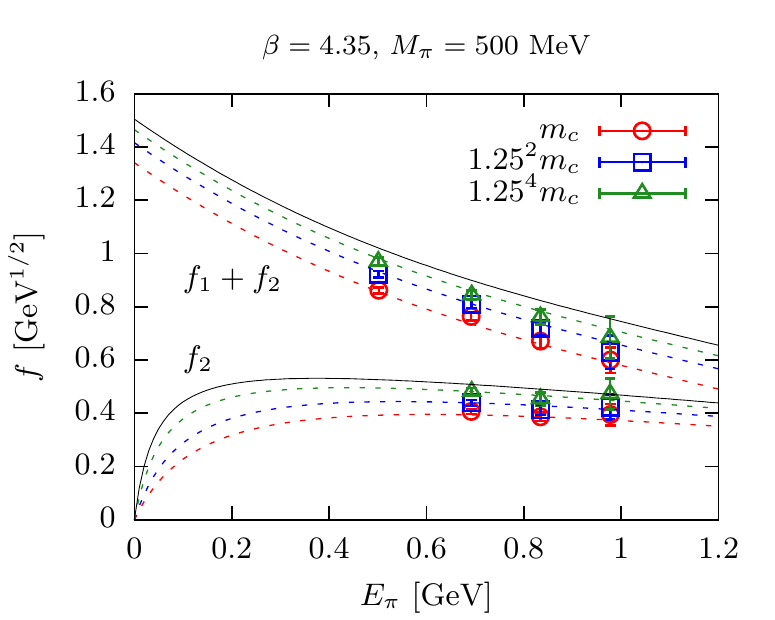}
\caption{On the left: Extrapolation to physical pion mass. Red squares, blue circles and green triangles
  show the lattice data points at different light quark masses ($M_\pi=500$, $400$ and $300$~MeV)
  at fixed lattice spacing, when the heavy quark mass is kept fixed. The corresponding dotted
  lines show the fits at those values of the parameters, and the solid black line shows the continuum
  and physical pion mass extrapolation at fixed heavy quark mass.
  On the right: Extrapolation to physical $B$ mass. Red squares, blue circles and green triangles show
  the lattice data points at different heavy quark masses at fixed lattice spacing when the light quark
  mass is kept fixed. The corresponding dotted lines show the fits at those values of the parameters,
  and the solid black line shows the continuum extrapolation and extrapolation to physical $B$ mass at
  fixed pion mass. In practice, we do all extrapolations (to continuum and to physical pion and $B$ meson
  masses) simultaneously.}
\label{fig:ml_mh_extrap}
\end{figure}

The extrapolations are illustrated in Fig.~\ref{fig:ml_mh_extrap}. In the fit we do all extrapolations
in one step, but we have examined each extrapolation individually by changing only one of the masses,
$m_l$ or $m_h$, or the lattice spacing $a$, while keeping the other parameters fixed. The extrapolations are
seen to be smooth, and they affect the form factors in different directions: the pion extrapolation approaches
the physical form factors from above, whereas the extrapolation towards the $b$ quark mass approaches the
physical form factors from below. The smallest pion mass used in this study is $\sim 225$ MeV and the
largest $B$ meson mass is $\sim 3.4$ GeV, so the extrapolations are sizeable but well under control.
The three lattice spacings give good control over the continuum extrapolation.

\begin{figure}[bt]
\centering
\includegraphics[width=0.65\textwidth]{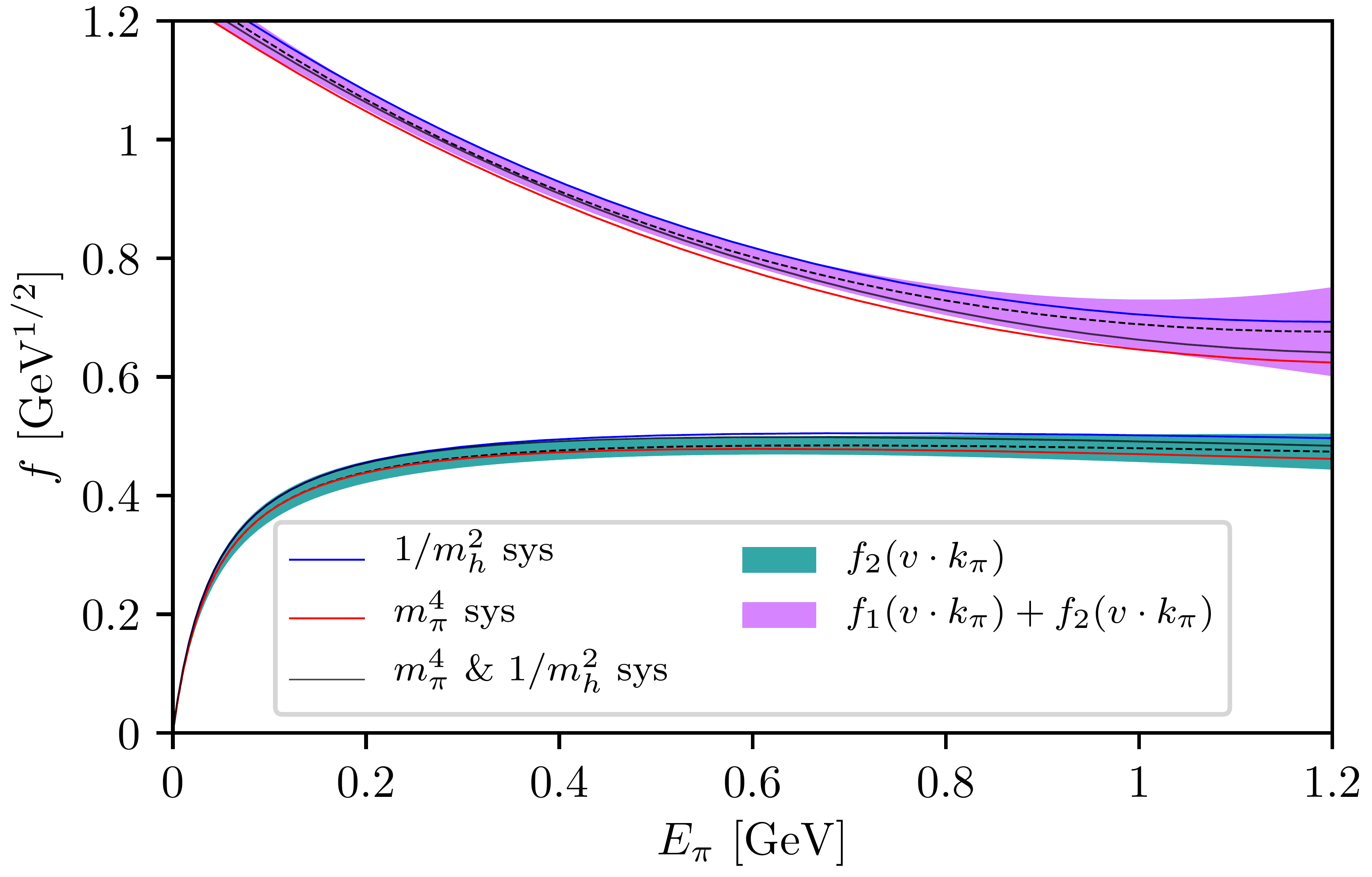}
\caption{Form factors $f_1(E_\pi)+f_2(E_\pi)$ and $f_2(E_\pi)$ in the continuum and physical
  limit. The dashed black line shows the original fit and the error bands show the statistical
  uncertainty. Blue, red and grey solid lines show the central values of the fits used in
  estimating systematic uncertainties by including higher order terms $1/m^2_h$, $M^4_\pi$,
  and both effects combined, respectively.
  }
\label{fig:ffsyst}
\end{figure}

We estimate the systematic effects in our result by re-doing the fit with higher order terms added
to the fit functions. Adding higher powers of $E_\pi$ changes the form factors $f_2(v\cdot k_\pi)$
and $f_1(v\cdot k_\pi)+f_2(v\cdot k_\pi)$ by 2-3\%. The effect of including $M_\pi^4$ terms is
roughly $-5$\%, whereas including terms proportional to $1/m^2_h$ increases the value of the form
factors by 5\%. This is illustrated in  Fig.~\ref{fig:ffsyst}. All in all, the systematic effects
are estimated to be roughly of the same size as the statistical uncertainty.

\section{$z$-expansion}

\begin{figure}[t]
\centering
\includegraphics[width=0.68\textwidth]{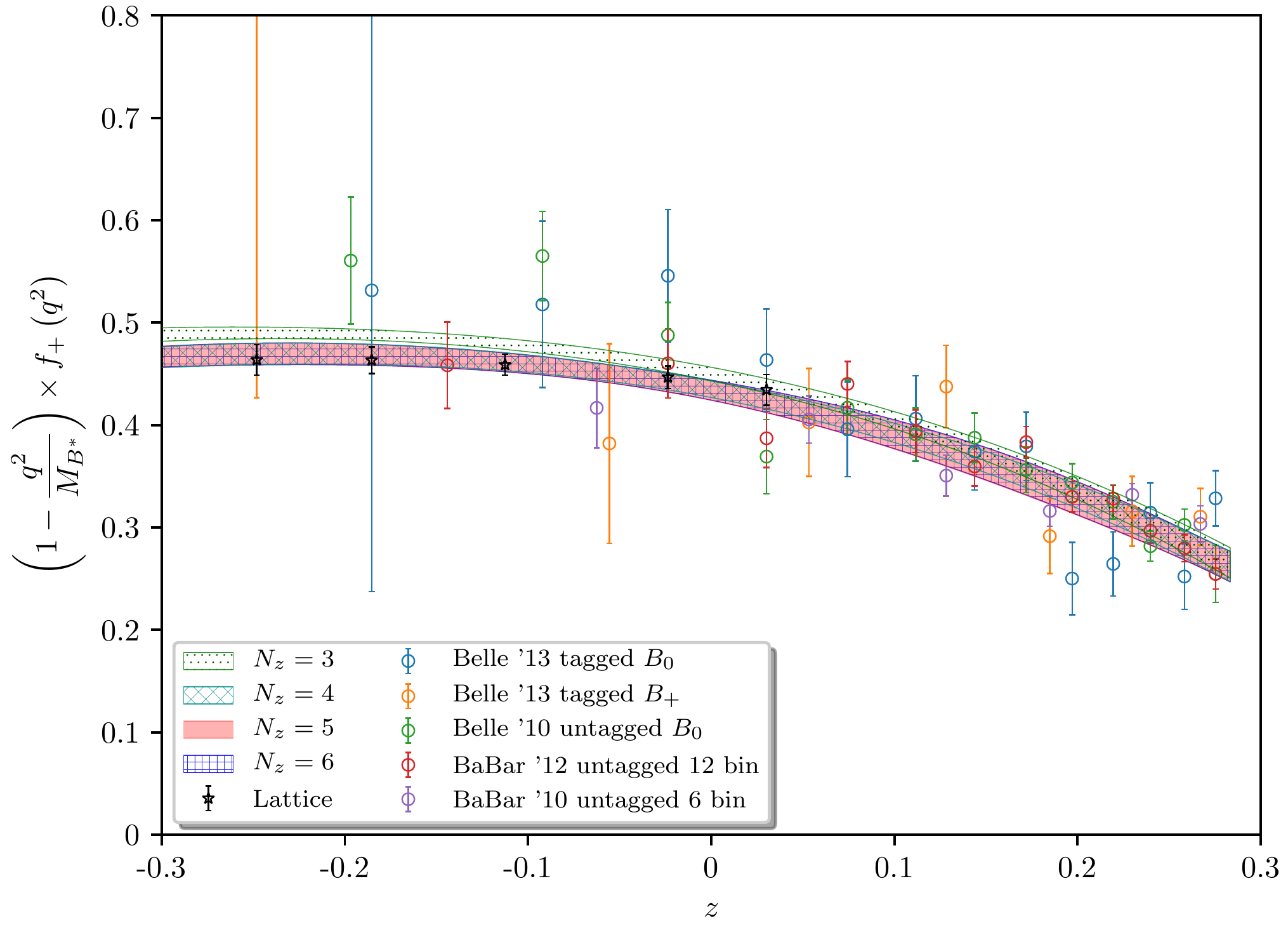}
\caption{z-expansion and combined fit to lattice and experimental data. The fit result for
  the form factor $(1-q^2/M^2_{B^\ast})f_+(q^2)$ in z-space with experimental and (synthetic) lattice data
  points.}
\label{fig:zexpansionfp}
\end{figure}

Experimental results of the differential decay rate for $B \to \pi\ell\nu$ are available from
both BaBar and Belle \cite{BaBar1,Belle1,BaBar2,Belle2}. Combining these experimental results
with our lattice calculation we can extract a value for the CKM matrix element $|V_{ub}|$. To do
this we pick synthetic data points from the lattice calculation of the form factors and fit them
together with the experimental data using the so-called $z$-expansion:
\begin{align}
\label{eq:zexpansion}
f_0(z)=&\sum_{n=0}^{N_z-1}a_nz^n, \\
f_+(z)=&\frac{1}{1-q^2(z)/M_{B^\ast}^2}\sum_{n=0}^{N_z-1}b_n\bigg[z^n-(-1)^{n-N_z}\frac{n}{N_z}z^{N_z}\bigg],\nonumber
\end{align}
where
\begin{equation}
z=\frac{\sqrt{t_+-q^2}-\sqrt{t_+-t_0}}{\sqrt{t_+-q^2}+\sqrt{t_+-t_0}},\quad t_+=(M_B+M_\pi)^2.
\end{equation}
By choosing $t_0=(M_B+M_\pi)(\sqrt{M_B}-\sqrt{M_\pi})^2$ we have a mapping between $q^2$ and $z$ where
the whole kinematic range is now $-0.3 < z < 0.3$. This is the advantage: $|z|$ is small and we can
use the expansion in powers of $z$ given in equation \eqref{eq:zexpansion}. We also utilize the
kinematic constraint $f_+(q^2=0) = f_0(q^2=0)$. $N_z = 5$ gives a good $\chi^2$, and adding higher
order terms does not change the result of the fit.

\begin{figure}[t]
\centering
\includegraphics[width=0.61\textwidth]{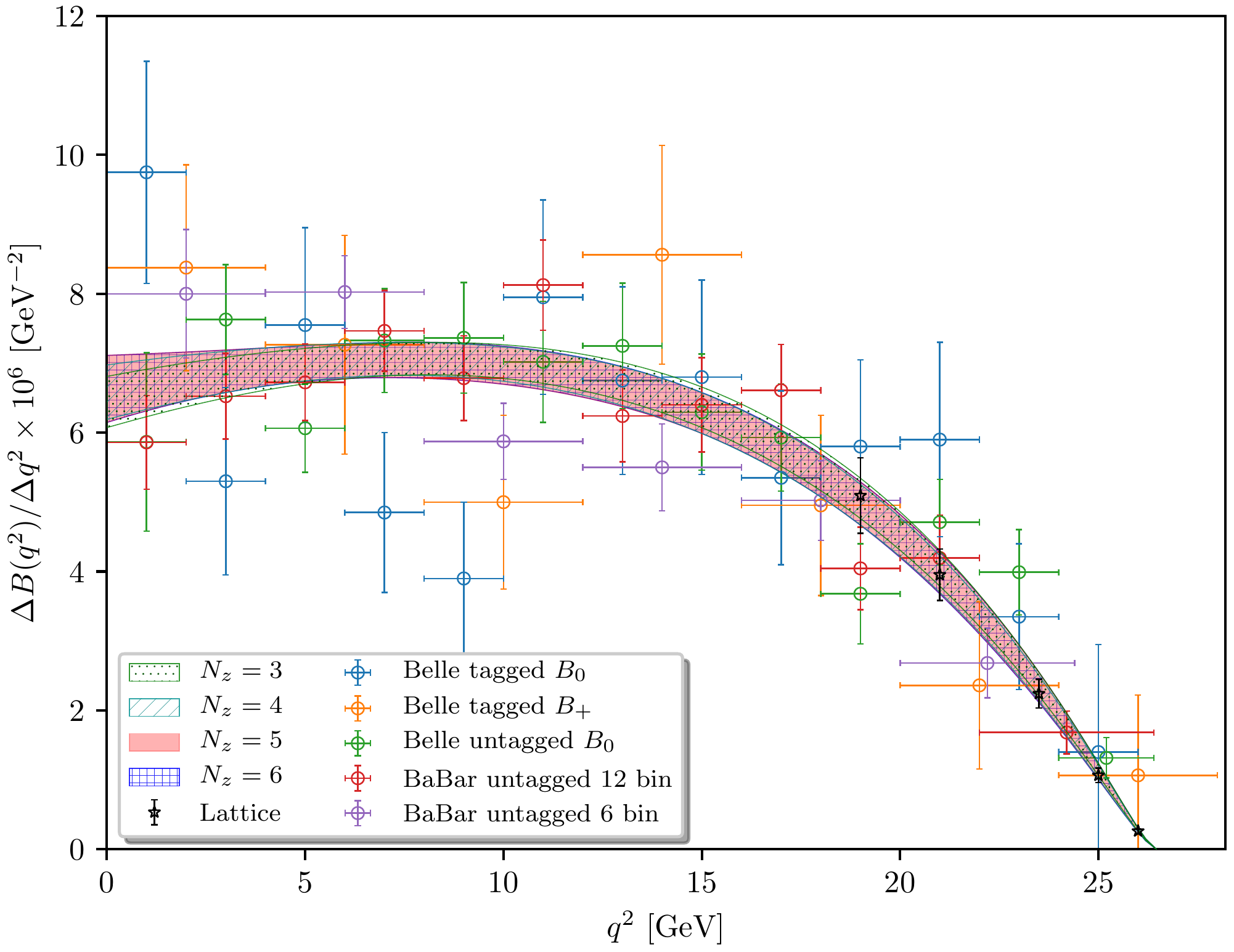}
\caption{z-expansion and combined fit to lattice and experimental data. The same fit and data points
  as in Fig.~\ref{fig:zexpansionfp}, but now in $q^2$ space. Here we present the data points and fit
  as branching fraction per $q^2$ bin instead of the form factor.}
\label{fig:zexpansiondeltaB}
\end{figure}

Equation \eqref{eq:decayrate} gives the relation between the differential decay rates and
the form factors, and $|V_{ub}|$ is included as a free fit parameter. Figures \ref{fig:zexpansionfp}
and \ref{fig:zexpansiondeltaB} summarize the results of the fit. The shape of the form factor from
experiment and lattice is in good agreement. Note that lattice results and experimental results are
highly complementary: lattice QCD results are available and most precise in the high $q^2$ region,
whereas experimental results are most precise in the low $q^2$ region.

\section{Conclusions}

Our results are still preliminary, as the combined fit to lattice and experimental data does not
contain all correlations, and full systematic errors are not included yet. This preliminary analysis
gives $|V_{ub}|=3.45(14)\times 10^{-3}$. The systematic uncertainties are likely to be of the same size
as the statistical uncertainties. 

\section*{Acknowledgements}

Numerical computations are performed on Oakforest-PACS at JCAHPC.
This work was supported in part by JSPS KAKENHI Grant Number JP18H03710 
and by MEXT as "Priority Issue on post-K computer".

\end{document}